\documentclass[preprint,superscriptaddress,amsmath,amssymb,prb]{revtex4-2}
\usepackage{graphicx} 
\usepackage{tabularx}
\usepackage{bm}

\begin{document}
\title{Variational Calculations of the Excited States of the Charged NV-center in Diamond Using a Hybrid Functional}

\author{Lei Sun}
 \affiliation{School of Physics, Peking University, Beijing 100871, People's Republic of China.}
\author{Elvar Örn Jónsson}
 \affiliation{Science Institute and Faculty of Physical Sciences, University of Iceland, 107 Reykjavík, Iceland}
\author{Aleksei Ivanov}
 \affiliation{Riverlane, St Andrews House, 59 St Andrews Street, Cambridge, CB2 3BZ, United Kingdom}
\author{Ji Chen}
 \email{ji.chen@pku.edu.cn}
 \affiliation{School of Physics, Peking University, Beijing 100871, People's Republic of China.}
 \affiliation{Interdisciplinary Institute of Light-Element Quantum Materials and Research Center for Light-Element Advanced Materials, Peking University, Beijing 100871, People's Republic of China}
 \affiliation{State Key Laboratory of Artificial Microstructure and Mesoscopic Physics and Frontiers Science Center for Nano-Optoelectronics, Peking University, Beijing 100871, People's Republic of China}
\author{Hannes Jónsson}
 \email{hj@hi.is}
 \affiliation{Science Institute and Faculty of Physical Sciences, University of Iceland, 107 Reykjavík, Iceland}

\begin{abstract}
The excited electronic states involved in the optical cycle preparation of a pure spin state of the negatively charged NV-defect in diamond are calculated using the HSE06 hybrid density functional and variational optimization of the orbitals. This includes the energy of the excited triplet as well as the two lowest singlet states with respect to the ground triplet state. In addition to the vertical excitation, the effect of structural relaxation is also estimated using analytical atomic forces. The lowering of the energy in the triplet excited state and the resulting zero-phonon line triplet excitation energy are both within 0.1 eV of the experimental estimates. An analogous relaxation in the lower energy singlet state using spin purified atomic forces is estimated to be 0.06 eV. These results, obtained with a hybrid density functional, improve on previously published results using local and semi-local functionals, which are known to underestimate the band gap. The good agreement with experimental estimates demonstrates how time-independent variational calculations of excited states using density functionals can give accurate results and, thereby, provide a powerful screening tool for identifying other defect systems as candidates for quantum technologies.
\end{abstract}

\maketitle


\section{Introduction}

Spin defects in solid-state materials have attracted significant attention in recent years due to their potential to enable quantum technologies operating near room temperature \cite{acin2018quantum}. 
Among them, the negatively charged nitrogen-vacancy (NV$^-$) center in diamond stands out as one of the most representative deep-level defects, offering unique advantages in quantum sensing, quantum communication, and quantum computing \cite{weber2010quantum,neumann2013high,barry2020sensitivity,dolde2013room,waldherr2014quantum}.
This prominence is largely attributed to its unique spin-optical properties, particularly the ability to initialize and read out its ground-state electron spin using optical fields, coupled with a remarkably long spin coherence time that persists under ambient conditions \cite{balasubramanian2009ultralong}.
The cornerstone of these capabilities is the optical spin polarization cycle. 
The system, initially in a degenerate spin-triplet ground state $^3A_2$, is optically excited to a corresponding triplet excited state $^3E$. 
Subsequently, a spin-selective, non-radiative decay pathway allows the system to preferentially relax back into the $m_s$ = 0 sub level of the ground state, thereby achieving a high degree of spin polarization \cite{weber2010quantum,dreyer2018first}.

Despite remarkable progress, accurately modeling the properties of such defects remains a critical challenge, particularly their excited states.
A variety of advanced computational methods have been employed \cite{kundu_quantum_2024,ma2010excited,chen2023multiconfigurational,ranjbar2011many,delaney2010spin,bockstedte2018ab,ma_quantum_2020,choi2012mechanism,huang2022simulating,chen2025towards} to calculate the electronic states for fixed atomic structure.
For instance, time-dependent density functional theory (TD-DFT) has been used to calculate vertical excitation energy values while considering quantum vibronic effects on the electronic energy levels \cite{kundu_quantum_2024}.
Other advanced approaches treating electron correlation at a higher level include the GW approximation and the Bethe-Salpeter equation \cite{ma2010excited}, parameterized effective Hubbard models \cite{ranjbar2011many,choi2012mechanism} and quantum embedding theories that solve a many-body Hamiltonian for the defect's active space using highly accurate full configuration interaction quantum Monte Carlo \cite{chen2023multiconfigurational,chen2025towards}.

These efforts have led to a growing consensus on the nature of the excited electronic states and the
energy involved in vertical excitations.
Another important challenge is the self-consistent treatment of the atomic structure, as the 
defect relaxes in different ways in the excited states.
While the aforementioned high-level methods provide an accurate description of the electronic wavefunctions, their prohibitive computational cost often prevents their direct use in calculations of full structural relaxations.
Consequently, most of these studies are on excitation energy for a fixed atomic structure, typically one that is optimized for the electronic ground state using a less computationally demanding method such as density functional theory (DFT) \cite{ma_quantum_2020} or based on an idealized crystal structure \cite{chen2025towards}. 
Such inconsistency between a high-level calculation of the electronic structure for an atomic structure obtained from a lower-level theory makes direct and accurate comparison with experimental values such as the zero phonon line (ZPL) difficult and complicates the validation of different theoretical approaches against one another \cite{doherty2013nitrogen}.

A powerful alternative approach are time-independent variational calculations using density functionals.
Different from conventional DFT calculations, convergence is reached on excited electronic states as higher-order stationary points on the electronic energy surface, thereby
providing solutions to the Kohn-Sham equations corresponding to higher energy than the ground state 
\cite{ivanov_electronic_2023,levi_variational_2020, schmerwitz2023calculations}.
In particular, the direct orbital optimization combined with the maximum overlap method (DO-MOM) has been shown to be successful \cite{ivanov_method_2021,levi_variational_2020,levi2020variational}.
It employs direct optimization to reduce computational cost and improve convergence rate \cite{Ivanov21b,levi_variational_2020,levi2020variational}. 
The maximum overlap method can be used to guide the 
calculation to a specific target excited state \cite{gilbert2008self},
or a projection of the gradient is used to converge on a saddle point of a give order in the generalized mode following (GMF) method 
\cite{schmerwitz2023calculations}.
This strategy has proven to be efficient for systems that pose notable challenges to other methods, including molecules with quasi-degenerate orbitals, charge-transfer excitations, and calculations of potential energy surfaces near conical intersections where conventional SCF-based methods \cite{van2002geometric,vandevondele2003efficient,baarman2011comparison} as well as linear response TD-DFT methods often fail \cite{dreuw2004failure,dreuw2005single,hait2021orbital}.
The variational nature of a solution corresponding to a stationary point on the electronic energy surface furthermore ensures compliance with the Hellmann-Feynman theorem, thereby granting access to nuclear forces for excited-state geometry optimizations and dynamics.
Recent application of DO-MOM to the NV$^-$ center using functionals within the local density approximation (LDA), generalized gradient approximation (GGA) and meta-GGA approximation has demonstrated promising results in that the ordering of the energy levels comes out to be right in each case, but the excitation energy has turned out to be strongly dependent on the chosen functional \cite{ivanov_electronic_2023}.
All these functionals underestimate the band gap of diamond and the energy of states within the gap can, therefore, also be expected to be underestimated.
It is clearly desirable to further study this system using a higher-level functional that gives band gap consistent with experiment and, furthermore, use it to assess structural relaxation in the excited electronic states.

In this work, we apply the DO-MOM method in combination with the HSE06 hybrid functional \cite{krukau2006influence, heyd2003hybrid}, which gives a band gap of diamond in close agreement with experimental data \cite{maciaszek2023application}, to study the vertical and adiabatic excitations of the NV$^-$ center.
Structural relaxations are calculated at the same level of theory as the estimation of the excitation energy.
Notably, we have 
used spin-purified atomic forces to optimize the structure of the $^1E$ singlet state.
The calculated ZPL results and the relaxation energy in the $^3E$ state are found to be in excellent agreement with available experimental estimates. 
We also analyze the atomic displacements of the $^1E$ and $^3E$ states and compare them with known Jahn-Teller distortions to further validate the qualitative nature of the structural
relaxations predicted by the calculations.


\section{Model and Methods}

\begin{figure}[h]
  \centering
  \includegraphics[width=1.0\textwidth]{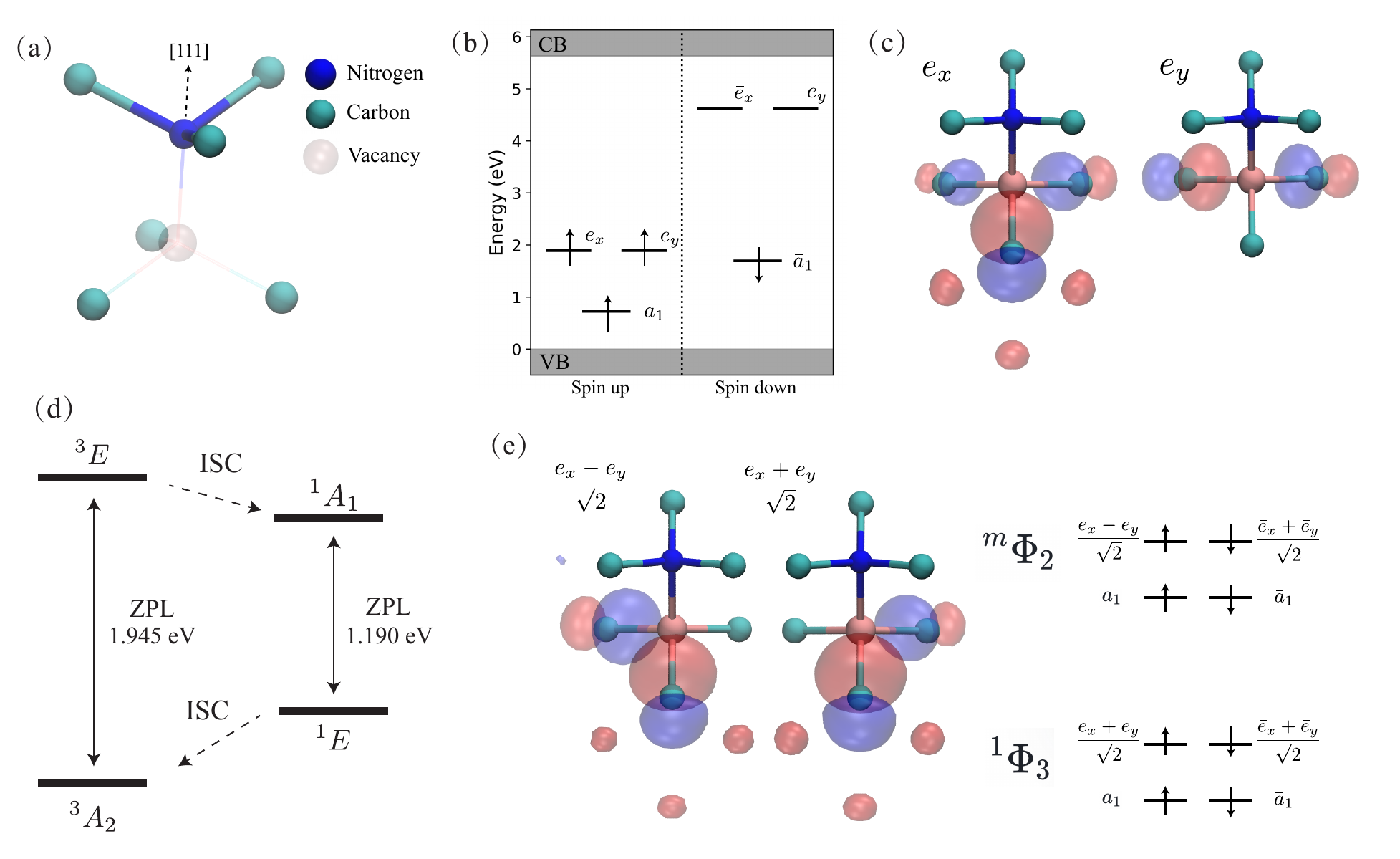}
  \caption{
  (a) Atomic structure of the NV$^-$ center and the threefold symmetry axis [111]. 
  (b) The energy levels of spin orbitals within the band gap. A bar over the orbital symbol indicates spin down. 
  (c) The orbitals $e_x$ and $e_y$. 
  (d) Schematic illustration of the optical cycle used for preparing the spin-pure ground state. 
  Measured values of the zero phonon lines (ZPL) of triplets \cite{davies_optical_1976} and singlets \cite{rogers2008infrared} are indicated. Intersystem crossings (ISC) are shown as dashed lines.
  (e) The highest occupied orbitals of mixed states $^m\Phi_2$ and $^1\Phi_3$. 
  Right hand side illustrate the occupied spin orbitals of $^m\Phi_2$ and $^1\Phi_3$.
  }
  \label{fig:fig1}
\end{figure}

The atomic structure of the NV$^-$ center consists of a substitutional nitrogen atom adjacent 
to a vacancy, as shown in Fig.~\ref{fig:fig1}(a). 
The system has three-fold symmetry axis and exhibits $C_{3v}$ point group symmetry.
Fig.~\ref{fig:fig1}(b) illustrates the spin orbitals associated with the defect and associated energy levels within the band gap for the ground triplet state, $^3A_2$.
The bar over an orbital symbol indicates spin-down channel.
The excited triplet state, $^3E$, is formed by exciting an electron from the $a_1$ orbital to one of the degenerate $e_x$ and $e_y$ orbitals, which are illustrated in Fig.~\ref{fig:fig1}(c).
The energy levels and the optical cycle are illustrated in Fig.~\ref{fig:fig1}(d) where the dashed lines indicate intersystem crossing (ISC) between singlet and triplet states.

The wave functions of singlet states are more complicated, but can be expressed 
as
\begin{align}
    & \Psi(^1\text{E}) = \left\{
    \begin{aligned}
    &(|e_x\bar{e}_y\rangle + |e_y\bar{e}_x\rangle) / \sqrt{2}\\
    &(|e_x\bar{e}_x\rangle - |e_y\bar{e}_y\rangle) / \sqrt{2}
    \end{aligned}
    \right.
    \label{eq:1e}
    \\ 
    & \Psi(^1\text{A}_1) = (|e_x\bar{e}_x\rangle + |e_y \bar{e}_y \rangle ) / \sqrt{2}
    \label{eq:1a1}
\end{align}
%
These states cannot be represented by a single Slater determinant \cite{lenef1996electronic,goss1996twelve}.
Hence, simply carrying out DFT calculations with same occupation number in both spin channels, the wavefunctions would converge to two symmetry broken mixed states, $^m\Phi_2$ and $^1\Phi_3$, which are linear combinations of pure spin states, $^m\Phi_2 = [\Psi(^1\text{E}) + \Psi(^3\text{A}_2)]/\sqrt{2}$, $^1\Phi_3 = [\Psi(^1\text{E}) + \Psi(^1\text{A}_1)]/\sqrt{2}$.
After spin purification, the energy of the pure singlet states can be expressed as 
\cite{ziegler1977calculation,von1979local}
\begin{align}
    & \varepsilon(^1\text{E}) = 2\cdot\varepsilon(^m\Phi_2) - \varepsilon(^3\text{A}_2)
    \label{eq:e_1e}
    \\
    & \varepsilon(^1\text{A}_1) = \varepsilon(^3\Phi_1) + 2 \cdot [\varepsilon(^1\Phi_3) - \varepsilon(^m\Phi_2)]
    \label{eq:e_1a1}
\end{align}

The force acting on atoms in the
spin-purified singlet states can be estimated by taking the partial derivatives of Eq.(\ref{eq:e_1e}) with respect to the atomic positions.
For example, the spin-purified force of the $^1E$ state is expressed in Eq.~(\ref{eq:f_1e}), where $f[^m\Phi_2]$ and $f[^3\text{A}_2]$ can be calculated using the standard Hellmann–Feynman procedure
\begin{align}
    f[^1\text{E}] = 2 \cdot f[^m\Phi_2] - f[^3\text{A}_2]
    \label{eq:f_1e}
\end{align}

In the DO-MOM procedure, the targeted $\bm{\Psi}$ is expressed as a unitary transformation $U$ of a set of auxiliary orbitals $\bm{\Phi}$, thus the total energy $\mathcal{E}(\bm{\Psi})$ is a function of $U$ and $\bm{\Phi}$
\begin{align}
 & \bm{\Psi} = U\bm{\Phi} \\
 & \mathcal{E}(\bm{\Psi}) = \mathcal{E}(U\bm{\Phi}) = \mathcal{F}(U,\bm{\Phi})
\end{align}

The stationary points for $\mathcal{E}(\bm{\Psi}$) can be found by two steps:

1. Find a stationary point of $\mathcal{F}(U,\bm{\Phi})$ with respect to $U$. 
Analytical first‑order partial derivative of $\mathcal{F}(U,\bm{\Phi})$ can be calculated using exponential transformation of the unitary matrix $U$ \cite{ivanov_method_2021}.
Limited-memory version of the symmetric rank-one quasi-Newton algorithm is used to optimized the parameters \cite{levi2020variational}.

2. Minimization of $\mathcal{F}(U,\bm{\Phi})$ where conventional energy function minimization algorithms generalized for a wave function optimization can be employed.
Detailed formalism and implementation of DO-MOM has previously been described  \cite{levi_variational_2020,ivanov_method_2021}.

The cubic supercell contains 215 atoms.
The optimized lattice constant of pure diamond at this level of theory is 3.546 \AA, giving a box length of 10.639 \AA.
Only the $\Gamma$ point of the Brillouin zone is included.
The projector augmented wave method is used in conjunction with a plane wave basis set with an energy cutoff at 600 eV.
The atomic coordinates are optimized until the atomic forces are all below 0.01 eV/\AA.
To achieve the symmetrized structure relaxation of the $^3E$ state, the occupation numbers of the HOMO and HOMO+1 orbitals are manually set to 1/2.
The calculations are carried out using the GPAW \cite{mortensen2024gpaw} software and the Atomic Simulation Environment suite \cite{larsen2017atomic}.


\section{Results and Discussions}

\begin{figure}[h]
  \centering
  \includegraphics[width = 0.8\textwidth]{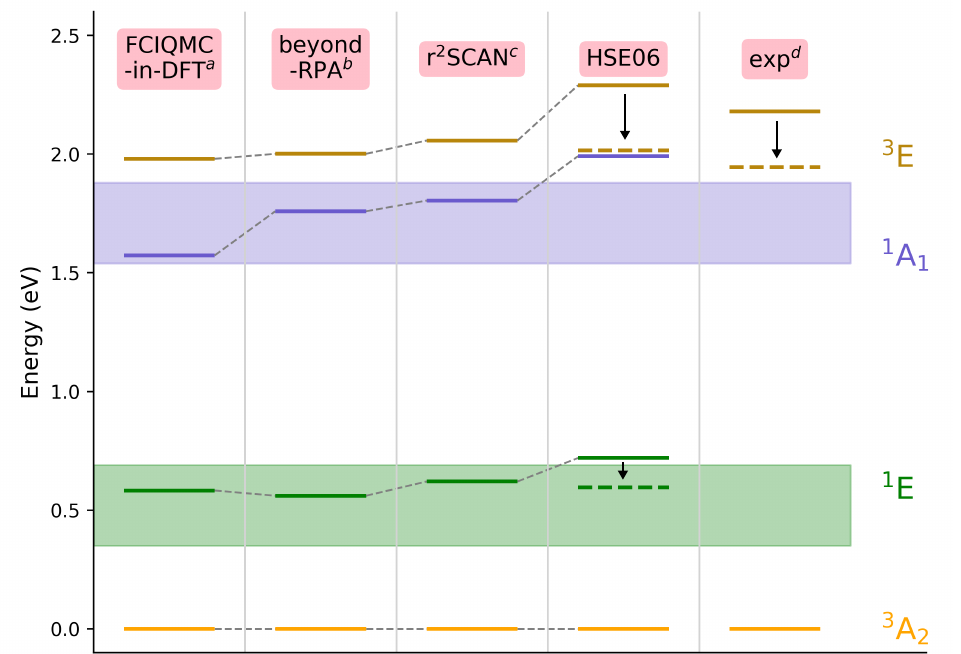}
  \caption{
  Energy levels of NV$^-$ center. 
  The solid lines denote vertical excitations and the dashed lines denote relaxed energy levels. 
  The colored region indicate the experimental estimates of $^1A_1$ and $^1E$ energy levels. 
  $^a$ Ref. \cite{chen2023multiconfigurational}, $^b$ Ref. \cite{ma_quantum_2020}, $^c$ Ref. \cite{ivanov_electronic_2023}, $^d$ Ref. \cite{davies_optical_1976,wolf_nitrogen-vacancy_2023}. Ref. \cite{davies_optical_1976} reports a vertical excitation energy of 2.180 eV and the ZPL of 1.945 eV. 
  Ref. \cite{wolf_nitrogen-vacancy_2023} gives the experimental estimates of the singlet states. 
  }
  \label{fig:energy-levels}
\end{figure}

The calculated relative energy of the four low-lying states of the NV$^-$ center is illustrated in Fig.~\ref{fig:energy-levels} and the values are listed in Table I.  
Our results are compared there against other theoretical approaches \cite{chen2023multiconfigurational,ma_quantum_2020,ivanov_electronic_2023} and reported experimental data \cite{davies_optical_1976, wolf_nitrogen-vacancy_2023}.
For clarity, we have only included theoretical data from three other works, including full configuration interaction quantum Monte Carlo with quantum embedding in DFT (FCIQMC-in-DFT) \cite{chen2023multiconfigurational}, beyond random phase approximation with quantum embedding in DFT (beyond-RPA) \cite{ma_quantum_2020}, and variational DFT calculations with the r$^2$SCAN exchange correlation functional  \cite{ivanov_electronic_2023}.
These data are good representatives of the current state-of-the-art.
The colored region in Fig. ~\ref{fig:energy-levels} indicates the experimental estimate of $^1A_1$ and $^1E$ energy levels \cite{wolf_nitrogen-vacancy_2023}.

For the transition from the $^3A_2$ ground state to the lowest-lying triplet excited state $^3E$, a transition that corresponding to the primary optical absorption feature of the NV$^-$ center, the HSE06 hybrid functional predicts a vertical excitation energy of 2.29 eV. 
This value can be compared with both experimental measurements and other theoretical results.
Experimentally, the absorption peak associated with this transition is well-established to be 2.180 eV \cite{davies_optical_1976}. 
Among the four theoretical values listed, FCIQMC-in-DFT, beyond-RPA, and r$^2$SCAN underestimate the results by 0.20, 0.18 and 0.12 eV, respectively.
We note, however, the reasons for these underestimations are different. 
For FCIQMC-in-DFT and beyond-RPA, while the treatments of electronic correlation are at high levels, their underestimations are most likely attributed to the limited size of the embedding cluster used in the calculations.
The r$^2$SCAN calculations are for a large periodic supercell, so there are negligible finite-size effects, but the meta-GGA nature of the r$^2$SCAN functional 
leads to an underestimate of the band gap and thereby likely also the position of the excited energy level 
in the gap.
Our HSE06 calculation addresses these limitations. 
It improves upon r$^2$SCAN by reducing the self-interaction error through the introduction of Fock exchange,
while also beying applicable to a large periodic supercell.

Following the vertical excitation calculation, a structural relaxation of the $^3E$ excited state was carried out to determine the minimum energy atomic structure.
This relaxation process accounts for the atomic rearrangements that occur after electronic excitation, as the excited-state electron density distribution often stabilizes a slightly different atomic configuration.
The relaxation lowers the energy to 2.01 eV.
This relaxed energy corresponds to the theoretical ZPL energy of the transition, i.e. the energy of the pure electronic transition without coupling to lattice phonons, which is the most reliable benchmark for comparison with the experimental ZPL.
This calculated ZPL energy deviates from the experimental value of 1.945 eV \cite{davies_optical_1976} by only 3\%, a level of agreement that is considered excellent for 
electronic structure calculations of defect centers in solids.
While both of the vertical and adiabatic energy are slightly overestimated in the HSE06 calculations, the relaxation energy of $^3E$ is estimated to be 0.28 eV, which is in good agreement to the experimental estimate of 0.235 eV \cite{davies_optical_1976}.

We now turn to the singlet excited states, which are essential for an understanding of the full optical 
processing via inter-system crossing. 
Experimentally, the energy of the lower energy singlet state, $^1E$, relative to the $^3A_2$ ground state has not been precisely determined as a single value but rather estimated to lie within a range of 0.35 to 0.69 eV, arising from the variability of different types of experiments.
When evaluating the vertical excitation energy, the HSE06 calculation predicts a value of 0.72 eV for the $^1E$ state, which is slightly above the upper bound of the experimental range. 
However, after performing structural relaxation of the $^1E$ state using the spin-purified forces, the energy of the $^1E$ state is lowered to 0.60 eV, falling comfortably within the experimentally reported range. 

The energy of the higher-lying singlet state, $^1A_1$, can be calculated using Eq.~(\ref{eq:e_1a1}).
The structure is chosen to be the symmetrized geometry of the $^3E$ state, because a limitation arises that the structural relaxation process based on the derived Hellmann-Feynman forces fails to preserve the symmetry of the $^1A_1$ state. 
Consequently, this structural relaxation would disrupt the $^1A_1$ state itself.
As shown in Fig.~\ref{fig:energy-levels}, our results show that the $^1A_1$ state lies 2.01 eV above the $^3A_2$ ground state. 
This value is slightly higher than the upper bound of the experimentally reported range, but that is not 
unexpected 
because of the choice of using the symmetrized geometry of the $^3E$ state in the calculation of the $^1A_1$ state.
There are three reasons for this choice.
First, the $^1A_1$ state is inherently symmetrized, and Eq.~(\ref{eq:e_1a1}) is also derived using representations of the $C_{3v}$ group. A symmetry broken atomic configuration could induce significant changes in orbitals and result in convergence problems.
Second, the $^3E$ state exhibits minimal distortion from the symmetrized configuration. The root mean square displacement of all atoms is only 0.07\AA, which is relatively small compared to the atomic displacements from $^3$A$^2$ to $^3E$ state listed in Table \ref{table:displacements}. The corresponding energy change in $^3E$ state is 30.0 meV, another small value when compared to the $^3E$ state's relaxation energy of 0.28 eV.
Third, in many experimental studies of the optical cycle, the $^1A_1$ is populated via transitions from the $^3E$ state. Additionally, the transition from $^1A_1$ to $^1E$ does not involve ISC.
As a result, the calculated energy of the $^1A_1$ can be considered to be in a relatively good agreement with the upper bound of experimentally reported values.
Structural relaxation would most likely to reduce its energy and possibly bring it within the experimental range; however, as discussed earlier, further work is required to develop a structural relaxation methodology that maintains the symmetry of the $^1A_1$ state.

\begin{table}[h]
\centering
\begin{tabularx}{0.9\textwidth}{p{4.0cm}XXX}
  \hline
   & $^1E$ / eV & $^1A_1$ / eV & $^3E$ / eV \\
  \hline
  FCIQMC-in-DFT$^a$ & 0.583 & 1.573 & 1.98 \\
  \hline
  beyond-RPA$^b$ & 0.56 & 1.76 & 2.00 \\
  \hline
  r$^2$SCAN$^c$ & 0.62 & 1.80 & 2.06 \\
  \hline
  HSE06 & 0.72 $\rightarrow$ 0.60 & 2.01 & 2.29 $\rightarrow$ 2.01 \\
  \hline
  exp.$^d$ & [0.35, 0.69] & [1.54, 1.88] & 2.180 $\rightarrow$ 1.945 \\
  \hline
\end{tabularx}
    \label{table:energy-levels}
\caption{
Various estimates of the energy levels of the NV$^-$ center, theoretical and experimental.
On the left side of the arrows are values of the vertical excitations energy, while on the right side are values of the adiabatic excitation energy.
FCIQMC-in-DFT stands for full configuration interaction quantum Monte Carlo with quantum embedding in DFT, beyond-RPA stands for beyond random phase approximation with quantum embedding in DFT, and r$^2$SCAN result is from variational DFT calculations with the r$^2$SCAN exchange correlation functional \cite{ivanov_electronic_2023}.
The numbers within brackets give the range in experimentally estimated values for the singlet states.
$^a$ Ref. \cite{chen2023multiconfigurational}, $^b$ Ref. \cite{ma_quantum_2020}, $^c$ Ref. \cite{ivanov_electronic_2023}, $^d$ Ref. \cite{davies_optical_1976,wolf_nitrogen-vacancy_2023}
}
\end{table}

\begin{figure}[h]
  \centering
  \includegraphics[width = 0.8\textwidth]{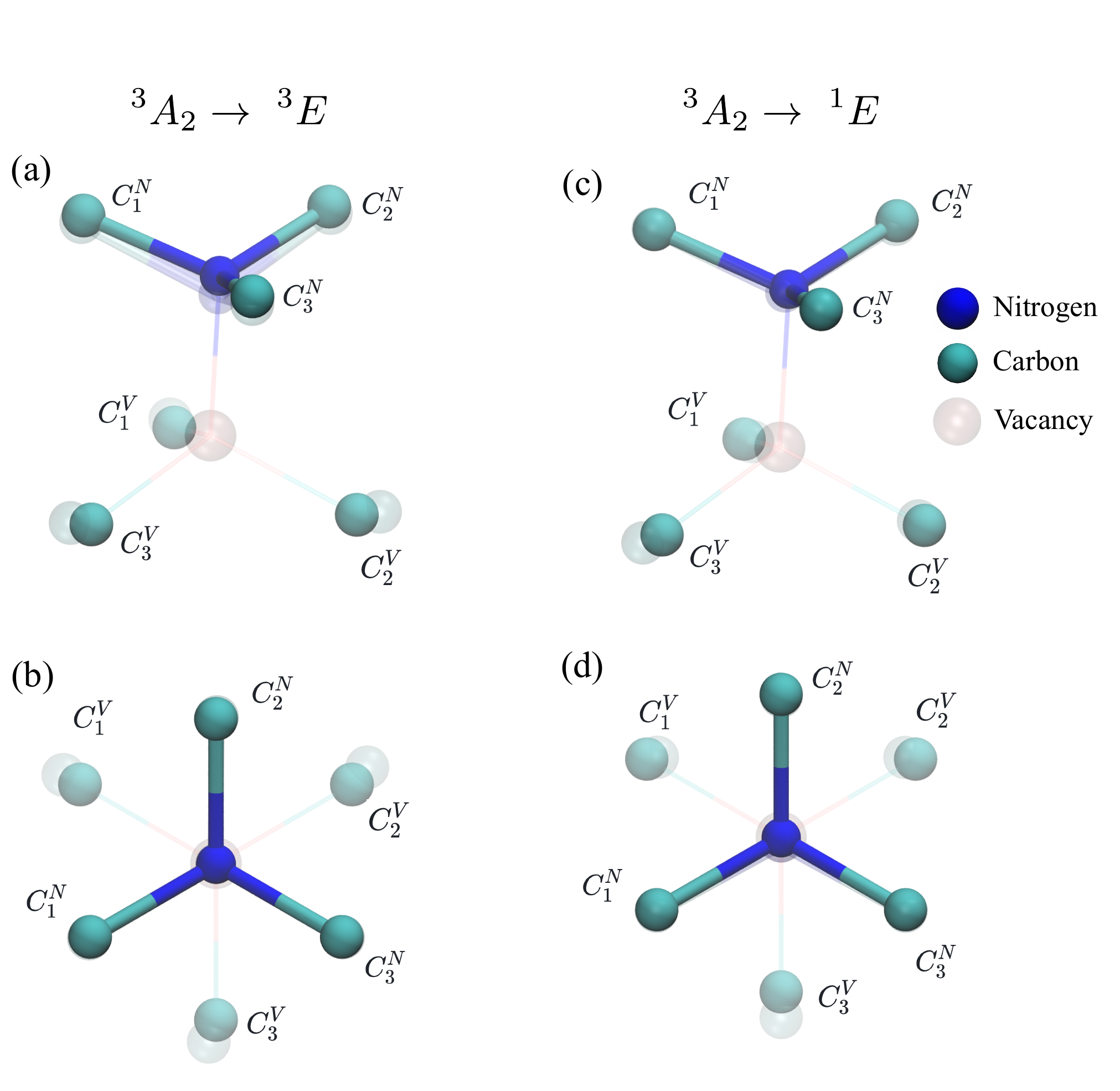}
  \caption{
  (a-b) Atomic displacements from $^3A_2$ to $^3E$ shown from two different viewpoints. The opaque and transparent atoms are from $^3A_2$ and $^3E$, respectively. (c-d) The atomic displacements from $^3A_2$ to $^1E$. The opaque and transparent atoms are from $^3A_2$ and $^1E$, respectively. 
  Structural relaxation is highly localized around the NV defect.
  Apart from the displaced atoms adjacent to the NV center, the atomic displacements of all other carbon atoms in the supercell are approximately one order of magnitude smaller and 
  are omitted from the figure for clarity.
  The displacements are doubled for clearer visualization.
  }
  \label{fig:displacements}
\end{figure}


\begin{table}[h]
\centering
\begin{tabularx}{0.9\textwidth}{p{3cm}XXXXXXXX}
  \hline
  & N & C$^\text{N}_1$ & C$^\text{N}_2$ & C$^\text{N}_3$ & C$^\text{V}_1$ & C$^\text{V}_2$ & C$^\text{V}_3$  \\ 
  \hline
  $^3A_2$ to $^3E$ /\AA & 0.066 & 0.024 & 0.024 & 0.026 & 0.062 & 0.076 & 0.076 \\
  \hline
  $^3A_2$ to $^1E$ /\AA & 0.019 & 0.005 & 0.012 & 0.005 & 0.035 & 0.035 & 0.084 \\
  \hline 
\end{tabularx}
  \label{table:displacements}
\caption{Displacements of atoms during energy minimization in the excited states.}
\end{table}

To further examine the relaxation effects, 
Fig.~\ref{fig:displacements} illustrates the atomic displacements from the $^3A_2$ state to the $^3E$ state and from the $^3A_2$ state to the $^1E$ state, respectively.
The diamond lattice exhibits full $C_{3v}$ symmetry, and the same applies to the $^3A_2$ and $^1A_1$ states.
In contrast, spatial symmetry is spontaneously broken in both the $^3E$ and $^1E$ states.
As illustrated in Fig.~\ref{fig:displacements} (a-b), for the relaxation process from $^3A_2$ to $^3E$ state, the nitrogen atom is pulled closer to the vacancy, along with three adjacent carbon atoms. 
Meanwhile, the carbon atoms near the vacancy are pushed outward, moving in the direction opposite to the nitrogen atom.
The displacements show strong $a_1$ character, while the anisotropy is rather subtle, indicating this relaxation is dominated by the intrinsic potential energy surface of the $^3E$ state without interacting with Jahn-Teller phonons \cite{abtew2011dynamic, doherty2013nitrogen}.
Fig.~\ref{fig:displacements}(c,d) show the atomic displacements between $^3A_2$ and $^1E$ state. 
The displacement pattern corresponds to a typical $e$-mode phonon, indicating that the structure relaxation between $^3A_2$ and $^1E$ are affected by Jahn-Teller effects more substantially.
The analyses of displacements further strengthen our conclusion that the HSE06 calculation with DO-MOM method can correctly optimize the structure of these excited states, paving the way for further theoretical analyses on Jahn-Teller and non-adiabatic effects in the future.


\section{Conclusion}

In summary, the HSE06 hybrid functional is used in variational DO-MOM calculations of the excited-states of the NV$^-$ center in diamond. 
Accurate estimates of the excitation energy are obtained for both excited triplet states and the lowest singlet state.
The vertical excitation energy from the $^3A_2$ ground state to the $^3E$ excited triplet state is determined to be 2.29 eV.
The relaxation energy is calculated to be 0.28 eV, and the resulting ZPL energy is then 2.01 eV. Both values are in close agreement with experimental measurements, with deviations within 0.1 eV.
These values of the excitation energy are higher than those previously obtained using local and semi-local 
functionals, consistent with the well-known underestimation of band gaps by such functionals, while the HSE06 predicts a band gap that closely matches experimental measurements.
By using spin-purified atomic forces, the optimized atomic structure of the 
$^1E$ singlet state is determined and the relaxation energy is estimated to be 0.06 eV.
The energy of the state is then 0.60 eV above the $^3A_2$ ground state, which aligns well with experimental estimates.
The analysis of atomic displacements also demonstrates that the structural relaxations obtained correctly capture the expected Jahn-Teller distortions in both $^3E$ and $^1E$ states, providing useful input and tools for deeper analyses of such effects using complementary theoretical frameworks that rely on the structures obtained with DFT.
Our work highlights the potential of time-independent variational DFT methods for future studies of excited states of solid state quantum systems.

\begin{acknowledgments}
This work was supported by the National Key R\&D Program of China under
Grant No. 2021YFA1400500, National Science Foundation of China under grant no. 12334003,
and by the Icelandic Research Fund under grant no. 2410644.
We thank the High Performance Computing Platform of Peking University for computational resources.
\end{acknowledgments}

\bibliography{ref.bib}

\end{document}